\documentclass[12pt,aps,floats,showpacs]{revtex4}
\usepackage{graphicx}
\usepackage{epsfig}

\newcommand{\be}{\begin{equation}}
\newcommand{\ee}{\end{equation}}
\newcommand{\bea}{\begin{eqnarray}}
\newcommand{\eea}{\end{eqnarray}}

\begin{document}
\title{Classical paths in systems of fermions}

\author{David H. Oaknin}

\affiliation{
Department of Physics and Astronomy, \\
University of British Columbia, Vancouver V6T 1Z1, Canada \\
email: doaknin@physics.ubc.ca}

\begin{abstract}

We implement in systems of fermions the formalism of pseudoclassical 
paths that we recently developed for systems of bosons and show that 
quantum states of fermionic fields can be described, in the Heisenberg 
picture, as linear combinations of randomly distributed paths that do not 
interfere between themselves and obey classical Dirac equations. Every 
physical observable is assigned a time-dependent value on each path in a 
way that respects the anticommutative algebra between quantum operators 
and we observe that these values on paths do not necessarily satisfy the 
usual algebraic relations between classical observables. We use these 
pseudoclassical paths to define the dynamics of quantum fluctuations in 
systems of fermions and show that, as we found for systems of bosons, the 
dynamics of fluctuations of a wide class of observables that we call 
"collective" observables can be approximately described in terms of 
classical stochastic concepts. Finally, we apply this formalism to 
describe the dynamics of local fluctuations of globally conserved fermion
numbers.

\end{abstract}

\maketitle

\section{Introduction}

Quantum fluctuations of physical observables can be manifest when they 
couple to another external system. For example, the Lamb shift 
in the absortion/emission lines of atoms is a direct consequence of their 
interaction with vacuum fluctuations of the electromagnetic fields. 
Similarly, quantum fluctuations in the energy-momentum tensor can affect 
the spacetime curvature and could result in dissipative effects 
in quantum cosmology and quantum gravity \cite{Hartle:nn,reynaud}.

The fluctuations and their dynamics can be described using the general 
formalism of time dependent Green functions to obtain the correlation of 
physical observables at different times. Some aspects of quantum 
fluctuations of commuting physical observables, like 
photon shot noise in optical systems or the Casimir force between plates, 
can be simulated as classical stochastic fluctuations 
with well defined probability distribution functions. However, it is not 
clear in this context to what extent it is possible to describe 
additional aspects of the dynamics of quantum fluctuations using 
classical stochastic concepts. 

In bosonic systems classicality is commonly associated with highly 
populated coherent states that can result, for example, from high 
temperatures or strong particle production in cosmology. In this context 
issues concerning classicality and decoherence of quantum fluctuations  
\cite{linde,polarski} during and after inflation, and their time 
evolution \cite{garriga,bucher} were addressed. Fermionic statistics, on 
the contrary, prevents any state to be highly populated and, moreover, 
fermionic coherent states cannot be identified with classical trajectories 
\cite{negele}, so it is generally conceded that there is no issue of 
classicality in quantum fluctuations of fermionic fields. 

In a recent paper \cite{oaknin} we have developed a new formalism of 
pseudoclassical incoherent paths to describe the dynamics of 
quantum fluctuations. New aspects of classical behaviour, in particular 
classical dynamics and decoherence, emerge in this formalism associated 
to fluctuations of collective observables which depend on a large number 
of degrees of freedom, even in scarcely populated states like the vacuum 
(a conclusion anticipated in \cite{Halliwell:1998jf} by different means). 
The notion of classicality that we propose is quite different than the 
case which is often discussed in the literature: 
it is associated to collective observables rather than to a certain class 
of quantum states, namely coherent states. In \cite{oaknin} we elaborated 
these ideas for a system of weakly interacting bosons. In this paper we 
show that this notion of classicality is not exclusive to systems of 
bosons, as it also emerges associated to fluctuations of collective 
observables in systems with a large number of fermions.
	
In the new formalism quantum states can be represented, in the Heisenberg 
picture, as linear combinations of randomly distributed pseudoclassical 
paths (PCP's) that do not interfere between themselves. We use each set 
of random PCP's to define the dynamics of quantum fluctuations in that 
state of the system. We remark that any single PCP cannot be identified 
with any quantum state. Instead, a whole set of PCP's with their 
corresponding probabilities to randomly happen at any time is the only 
formal object that can describe the dynamics of quantum fluctuations in 
one state of the system. On each path of a set every physical observable 
is assigned a time-dependent value in a way that is consistent with the 
commutation/anticommutation relations between quantum operators. In 
particular, the generalized canonical coordinates and their conjugate 
momentum operators get time-dependent values which obey classical 
equations of motion and depict a collection of harmonic oscillators with 
constrained initial conditions. On the other hand, we notice that these 
pseudoclassical time-dependent values on paths do not necessarily respect 
the usual algebraic relations that classical observables fulfill, as a 
consequence of the non-commutative relations between quantum operators. 

We select collective observables which depend on a large number of 
bosonic or fermionic independent degrees of freedom because we 
realize that their time-dependent values on paths do approximately 
regain the usual algebraic relations between classical observables and, 
therefore, we can use the formalism of PCP's to describe the dynamics of 
quantum fluctuations of collective observables in terms of unconstrained 
classical stochastic processes. 

The formalism of PCP's, for bosons or fermions, shows that we can give 
a description of quantum mechanical states in terms of non-interfering
paths if we trade it for a non-trivial definition of the algebraic 
relations between the values on paths of physical observables. This 
formalism can help to understand the process of decoherence and the 
onset of classicality in quantum systems because it can be applied to 
closed systems without explicit reference to observers, measurement, or 
an environment. In some cases it may also become a useful tool for 
performing calculations, in particular numerical simulations. 

This paper is organized in five sections. In section II we review 
the fundamental concepts of the formalism of PCP's and implement them in a 
system with a single fermion mode. In section III we extend the formalism 
to a system with two non-interacting fermions, and then we generalize it 
to systems with any finite number of free fermions. In section IV we 
define collective observables and describe their dynamics in the context of a 
regularized QFT. Section V contains a summary of results. This paper deals
basically with linear fermionic fields. A discussion of the general case
of interacting fields is postponed to a forthcoming work.

\section{Pseudoclassiclal paths of a single fermionic mode}

In this section we present a detailed description of a system with a 
single fermionic mode in the new formalism of PCP's. We will
start building the set of random paths which describes the dynamics of
quantum fluctuations in the vacuum state and, afterwards, will show how to 
build the sets of paths that describe states other than the vacuum.
 
The hamiltonian of the system is
\begin{equation}
\label{hamiltonian}
H = \kappa \left(a^{\dagger}a - \frac{1}{2}\right),
\end{equation}
where the operators $a^{\dagger}$, $a$, which create and annihilate
the excitations of the mode, obey anticommutation relations
$\{a^{\dagger},a\}=1$ and $\{a^{\dagger},a^{\dagger}\}=\{a,a\}=0$.
The last two relations imply, in particular, that
$(a^{\dagger})^2=a^2=0$ and, therefore, the orthonormal basis of   
eigenstates of the hamiltonian, which linearly span the Hilbert space of 
the states of the system, contains only two independent vectors: 
$\{|0 \rangle, |1 \rangle\}$. The action 
of the operators $a^{\dagger}$, $a$ on this basis is given by the equations
$\label{destruction} a|0 \rangle=0$, $a|1 \rangle=|0 \rangle$
and $\label{creation} a^{\dagger}|0 \rangle=|1 \rangle$, 
$a^{\dagger}|1 \rangle=0$. We will fix for simplicity $\kappa=1$ 
and, therefore, $\label{basiseigenstates}
H|0 \rangle = -\frac{1}{2}|0 \rangle$ and 
$H|1 \rangle = +\frac{1}{2}|1 \rangle$. 

The operator $a^{\dagger}$ is the hermitic conjugate of the operator 
$a$, so that the couple of operators
\begin{equation}
\label{hermitic}
\xi=\frac{1}{\sqrt{2}}(a+a^{\dagger}), \hspace{1.0in}
{\widetilde \xi}=\frac{i}{\sqrt{2}}(a-a^{\dagger}), 
\end{equation}
are hermitic and obey anticommutation relations:
$\label{anticommutation}
\{\xi,\xi\}=\{{\widetilde \xi},{\widetilde \xi}\}=1$
and $\{{\widetilde \xi},\xi\}=0$.

The action of these operators in the basis of 
eigenstates of the hamiltonian is summarized in the equations
$\xi|0 \rangle = \frac{1}{\sqrt{2}}|1 \rangle$, 
$\xi|1 \rangle = \frac{1}{\sqrt{2}}|0 \rangle$ and
${\widetilde \xi}|0 \rangle =-\frac{i}{\sqrt{2}}|1 \rangle$, 
${\widetilde \xi}|1 \rangle = \frac{i}{\sqrt{2}}|0 \rangle$,
and, therefore, they can be identified with Pauli matrices, 
$\xi=\frac{1}{\sqrt{2}} \sigma_1$ and
${\widetilde \xi}=-\frac{1}{\sqrt{2}} \sigma_2$. Notice that
$\xi^2={\widetilde \xi}^2=\frac{1}{2}$.

The most general linear operator that can be defined on the
Hilbert space of the single fermionic mode is a polynom of the type,
$\label{grassmann}
O = \alpha_1 \cdot 1 + \alpha_2 \cdot \xi + \alpha_3 \cdot {\widetilde
\xi} + \alpha_4 \cdot i{\widetilde \xi}\xi$,
where the coefficients $\alpha_j, j=1,2,3,4$ are, in general, 
complex numbers. This linear set of operators is the Grassmann algebra 
generated by the anticommuting hermitic operators $\xi$,${\widetilde 
\xi}$. The operator $\xi$ (or independently, ${\widetilde \xi}$) forms a 
complete representation of commuting observables in the Hilbert space.
An operator $O$ is hermitic if and only if the 
coefficients $\alpha_j, j=1,2,3,4$ are all real. In particular, the 
hamiltonian (\ref{hamiltonian}) is
\begin{equation}
\label{newhamiltonian}
H= i {\widetilde \xi} \xi
\end{equation}

Our first aim is now to write the action of a generic operator
$O$ on the vacuum state in terms of the action on the same state of some 
linear combination of the identity operator $1$ and the generator $\xi$.
In general, it can be immediately checked that
$O|0 \rangle = \left((\alpha_1-\frac{1}{2}\alpha_4)\cdot 1 + 
(\alpha_2-i\alpha_3)\xi \right)|0 \rangle
\equiv {\cal P}_o(1,\xi)|0 \rangle$. It is also immediate to check 
that the operator ${\cal P}_o(1,\xi)$ defined through this equation is
unique as the identity $\beta_1 \cdot 1|0 \rangle=\beta_2 \cdot \xi|0 
\rangle$ only holds if $\beta_1=\beta_2=0$. From the identities, 
\begin{equation}
\label{keyrelation}
{\widetilde \xi}|0 \rangle = -i\xi|0 \rangle, \hspace{0.8in} 
H|0 \rangle = -\frac{1}{2}|0 \rangle,
\end{equation}
we obtain, for example, that ${\cal P}_{\small{\widetilde 
\xi}}(1,\xi)=-i\xi$ and ${\cal P}_h(1,\xi)=-\frac{1}{2}\cdot 1$.

The next stage in our programme is to identify the two eigenstates of the 
operator $\xi$:
$\label{xirepresentationeigenvectors}
|q_+ \rangle = \frac{1}{\sqrt{2}}(|0 \rangle + |1 \rangle)$ and 
$|q_- \rangle = \frac{1}{\sqrt{2}}(|0 \rangle - |1 \rangle)$,
and their corresponding eigenvalues, 
$\lambda_+=\frac{+1}{\sqrt{2}}$ and $\lambda_-=\frac{-1}{\sqrt{2}}$. Then, 
we expand the vacuum state in the new basis: $|0 \rangle = 
\frac{1}{\sqrt{2}}(|q_+ \rangle + |q_- \rangle) \equiv \sum_{q=q_{\pm}} 
\psi(q) |q \rangle$, with $\psi(q_{\pm})=\frac{1}{\sqrt{2}}$.
The wavefunction of the vacuum in the basis of 
eigenstates of $\xi$, $\psi(q)$, is equally valued on 
each of the two disconnected points ${|q_+ \rangle,|q_- \rangle}$
of the configuration space and so the probability 
$|\psi(q_{\pm})|^2=(\frac{1}{\sqrt{2}})^2=\frac{1}{2}$ of each of
them to happen randomly. 

Moreover, using the relation $\xi|0 \rangle =
\frac{1}{\sqrt{2}}(\xi|q_+ \rangle + \xi|q_- \rangle) =
\frac{1}{\sqrt{2}}(\lambda_+|q_+ \rangle + \lambda_-|q_- \rangle)$
we can assign to the operator $\xi$ the classical 
value
\begin{equation}
\label{xiplus}
\xi_{cl}(q_+)=\lambda_+=\frac{\langle q_+|\xi|0 \rangle}{\langle q_+|0 
\rangle}
=\frac{+1}{\sqrt{2}}
\end{equation}
on the first point, and
\begin{equation}
\label{ximinus}
\xi_{cl}(q_-)=\lambda_-=\frac{\langle q_-|\xi|0 \rangle}{\langle q_-|0 
\rangle}
=\frac{-1}{\sqrt{2}}
\end{equation}
on the second point of the configuration space. Thus, we have defined on 
configuration space a random variable $\xi_{cl}(q)$.

Furthermore, following the identity $O|0 \rangle={\cal P}_o(1,\xi)|0 
\rangle=\frac{1}{\sqrt{2}}({\cal P}_o(1,\xi_{cl}(q_+))|q_+ \rangle + 
{\cal P}_o(1,\xi_{cl}(q_-))|q_- \rangle)$, the generic operator $O$ should 
be 
assigned the random variable \begin{equation}
O_{cl}(q_{\pm})={\cal P}_o(1,\xi_{cl}(q_{\pm}))=
\frac{\langle q_{\pm}|{\cal P}_o(1,\xi)|0 \rangle}{\langle q_{\pm}|0 
\rangle},
\end{equation}
that gives the operator a pseudoclassical value at each point $|q_{\pm} 
\rangle$ of the vacuum configuration space. For example, looking at 
(\ref{keyrelation}) we assign to the operator ${\widetilde \xi}$ the 
random variable ${\widetilde \xi}_{cl}(q)$ which takes the value
\begin{equation}
\label{wxiplus}
{\widetilde \xi}_{cl}(q_+)=-i \xi_{cl}(q_+)=\frac{-i}{\sqrt{2}}
\end{equation}
on the first point of the configuration space and
\begin{equation}
\label{wximinus}
{\widetilde \xi}_{cl}(q_-)=-i \xi_{cl}(q_-)=\frac{i}{\sqrt{2}},
\end{equation}
on the second point. Also according to (\ref{keyrelation}), the free 
hamiltonian (\ref{newhamiltonian}) is 
assigned on each of these 
points the constant value
\begin{equation}
\label{classhamil}
h_{cl}(q_{\pm})=-\frac{1}{2}=\langle 0|H|0 \rangle.
\end{equation}

Let then go a step further to describe the time dependence of the random 
variables we have just defined. In the Heisenberg picture, time dependence 
of the operator $O$ is described by the expression $O(t)=e^{iHt} O 
e^{-iHt}$, where $H$ is the hamiltonian of the system. The observable 
$O(t)$ can then be associated following the same algorithm that we have 
described above with a new random variable $O_{cl}(t,q_{\pm})$. Thus, 
each physical observable is given a time-dependent c-value at each point 
$|q \rangle$ of the configuration space, that is a path whose realization 
probability is $|\psi(q)|^2$.  

In particular, the time evolution of the operators $\xi$ and 
${\widetilde \xi}$ is described by the expressions
\begin{equation}
\label{xip}
\xi(t) = e^{iHt} \xi e^{-iHt} = cos(t) \xi - sin(t) {\widetilde \xi}
\end{equation}
\begin{equation}
\label{xim}
{\widetilde \xi}(t) = e^{iHt} {\widetilde \xi} e^{-iHt} = sin(t) \xi +
cos(t) {\widetilde \xi},
\end{equation}
which solve the differential equations 
\begin{equation}
\label{fermiondynamics}
\frac{d\xi(t)}{dt} = -{\widetilde \xi}(t),
\hspace{1.0in}
\frac{d{\widetilde \xi}(t)}{dt} = \xi(t).
\end{equation}

On the vacuum state $|0 \rangle$ these equations imply 
(see (\ref{keyrelation}))
\begin{equation}
\xi(t)|0 \rangle = e^{it} \xi|0 \rangle,
\hspace{1.0in}
{\widetilde \xi}(t)|0 \rangle =-i e^{it} \xi|0 \rangle,
\end{equation}
which, according to our formalism, mean that in the ground state $|0 
\rangle$ the time-dependent classical values of this pair of operators on 
each of the two paths $|q_{\pm} \rangle$ are 
$\xi_{cl}(t,q_{\pm})=e^{it} \xi_{cl}(q_{\pm})$ and
${\widetilde \xi}_{cl}(t,q_{\pm})=-i e^{it} \xi_{cl}(q_{\pm})$. They
obey the classical equations of an harmonic oscillation
\begin{equation}
\label{classicalfermions}
\frac{d\xi_{cl}(t;q_{\pm})}{dt} = -{\widetilde \xi}_{cl}(t;q_{\pm}),
\hspace{1.0in}
\frac{{d\widetilde \xi}_{cl}(t;q_{\pm})}{dt} = \xi_{cl}(t;q_{\pm}),
\end{equation}
with initial conditions fixed by (\ref{xiplus} and \ref{wxiplus}) on the 
first path, and (\ref{ximinus} and \ref{wximinus}) on the second path.
The operator $i{\widetilde \xi}\xi$ commutes with the hamiltonian so
it does not change with time and, therefore, neither its classical value 
(\ref{classhamil}) on each path does. 

The formalism of PCP's can be extended to describe any other state in the 
Hilbert space of the single fermionic mode. Consider, for example, the 
normalized state $|\Psi \rangle= cos(\theta)|0 
\rangle+sin(\theta)e^{i\phi}|1 \rangle$ and let expand it in the basis
$\{|q_{\pm} \rangle\}$ of eigenstates of the generator $\xi$: 
\begin{equation}
\label{statePsi}
|\Psi \rangle=
\frac{1}{\sqrt{2}}\left((cos(\theta)+sin(\theta)e^{i\phi})|q_+
\rangle+(cos(\theta)-sin(\theta)e^{i\phi})|q_- \rangle
\right).
\end{equation}

On each of the two points $|q_{\pm} \rangle$ in the configuration space 
the random variable $\xi_{cl}(q_{\pm})$ takes the value $\frac{\langle 
q_{\pm}|\xi|\Psi \rangle}{\langle q_{\pm}|\Psi \rangle}=
\frac{\pm 1}{\sqrt{2}}$, identical to those specified by eq. 
(\ref{xiplus}) and (\ref{ximinus}). In order to obtain in the new quantum 
state $|\Psi \rangle$ the random variable which corresponds to the generic 
operator $O$ we must find in the linear subspace spanned by the identity 
operator $1$ and the generator $\xi$ the operator ${\cal P}^{\psi}_o(1,\xi)$ 
such that 
\begin{equation}
\label{ground}
O|\Psi \rangle = {\cal P}^{\psi}_o(1,\xi)|\Psi \rangle.
\end{equation}
This can be done through the following three steps:
1) we write the state $|\Psi \rangle$ as a linear combination of the 
operators $1$ and $\xi$ acting on the vacuum: $|\Psi \rangle = 
\left(cos(\theta) \cdot 1 +\sqrt{2}sin(\theta)e^{i\phi} \xi \right)|0 
\rangle$; 2) now, use the rules that we have stated before 
in this section in order to obtain the action of the operator $O$ on the 
state $|\Psi \rangle$ in terms of the action of some new linear 
combination of the operators $1$ and $\xi$ on the vacuum state:
$O|\Psi \rangle = O\left(cos(\theta)+ \sqrt{2}sin(\theta)e^{i\phi} \xi 
\right)|0 \rangle = {\cal G}(1,\xi)|0 \rangle$;
3) note that the relation introduced in step 1) can be inverted into
the relation $|0 \rangle = \frac{1}{cos^2(\theta)-sin^2(\theta)e^{2i\phi}}
\left(cos(\theta)-\sqrt{2}sin(\theta)e^{i\phi} \xi \right)|\Psi \rangle$, 
which is then introduced in the expression that we obtained from step 2) 
in order to get, as desired, the action of operator $O$ on state $|\Psi 
\rangle$ in terms of the action of some linear combination ${\cal 
P}^{\psi}_o(1,\xi)$ of the operators $1$ and $\xi$ on such 
quantum state.
The three described steps of this algorithm are well and uniquely defined, 
so the operator ${\cal P}^{\psi}_o(1,\xi)$ is unique for each operator 
$O$.

In particular, for the operator ${\widetilde \xi}$ we obtain the identity
\begin{equation}
\label{keyPsi}
{\widetilde \xi}|\Psi \rangle =
\frac{i}{cos^2(\theta)-sin^2(\theta)e^{2i\phi}}
(\sqrt{2}sin(\theta)cos(\theta)e^{i\phi} -
(cos^2(\theta)+sin^2(\theta)e^{2i\phi})\xi)|\Psi\rangle,
\end{equation}
which means that this operator is assigned, when the system is in the 
quantum state $|\Psi \rangle$, the random variable 
${\widetilde \xi}(q_{\pm})=\frac{\langle q_{\pm}|{\widetilde \xi}|\Psi 
\rangle}{\langle q_{\pm}|\Psi \rangle}$ whose value at each point of the 
configuration space is:
\begin{equation}
\label{psi}
{\widetilde \xi}_{cl}(q_{\pm}) =
\frac{i}{cos^2(\theta)-sin^2(\theta)e^{2i\phi}}
(\sqrt{2}sin(\theta)cos(\theta)e^{i\phi} -
(cos^2(\theta)+sin^2(\theta)e^{2i\phi})\xi_{cl}(q_{\pm})).
\end{equation}

The dynamical equations (\ref{fermiondynamics}), when 
applied on the quantum state $|\Psi \rangle$,
\begin{equation}
\label{dynamicalequations}
\frac{d\xi(t)|\Psi \rangle}{dt} = -{\widetilde \xi}(t)|\Psi \rangle,
\hspace{1.0in}
\frac{d{\widetilde \xi}(t)|\Psi \rangle}{dt} = \xi(t)|\Psi \rangle,
\end{equation}
enforce that classical values of the operators $\xi$ and ${\widetilde 
\xi}$ on each pseudoclassical path of the random set that describes the 
state $|\Psi \rangle$ still evolve according to classical equations of 
motion (\ref{classicalfermions}). Initial conditions are now fixed on each 
path by (\ref{xiplus}) and (\ref{ximinus}), respectively, for $\xi_{cl}(t=0)$ 
and (\ref{psi}) for ${\widetilde \xi}_{cl}(t=0)$, instead of  
(\ref{wxiplus}), (\ref{wximinus}). In general, the state of the system
$|\Psi \rangle$ fixes only the initial conditions for its set of PCP's, 
while the dynamics of these paths subsequently follows classical 
equations. 

The action of the hamiltonian (\ref{newhamiltonian}) on the quantum state 
$|\Psi \rangle$ can be expressed, using (\ref{keyPsi}), as
\begin{equation}
H|\psi \rangle=-i\xi{\widetilde \xi}|\Psi \rangle=
\frac{1}{cos^2(\theta)-sin^2(\theta)e^{2i\phi}}
(\sqrt{2}sin(\theta)cos(\theta)e^{i\phi}\xi -
(cos^2(\theta)+sin^2(\theta)e^{2i\phi})/2)|\Psi\rangle,
\end{equation}
which implies that the two paths of the set that describes quantum 
fluctuations in the generic state $|\Psi \rangle$ do not necessarily have 
the same energy, but
\begin{equation}
\label{pce}
h_{cl}(q_{\pm}) =
\frac{1}{cos^2(\theta)-sin^2(\theta)e^{2i\phi}}
(\sqrt{2}sin(\theta)cos(\theta)e^{i\phi}\xi_{cl}(q_{\pm}) -
(cos^2(\theta)+sin^2(\theta)e^{2i\phi})/2).
\end{equation}

The two pseudoclassical paths either are no longer equally probable as can 
be seen from equation (\ref{statePsi}): when the system is in the quantum state 
$|\Psi \rangle$ the probability of the point $|q_+ \rangle$ to happen 
randomly is  
$|\psi(q_+)|^2=\frac{1}{2}|cos(\theta)+sin(\theta)e^{i\phi}|^2$, 
while the probability of the point $|q_- \rangle$ is 
$|\psi(q_-)|^2=\frac{1}{2}|cos(\theta)-sin(\theta)e^{i\phi}|^2$. 

Notice that although expressions (\ref{psi}) and (\ref{pce}) are 
ill-defined at the event $|q_- \rangle$ when $e^{i\phi}=\pm 1$ 
and $sin(\theta)=\pm cos(\theta)$, the probability of such event to happen 
randomly is zero, while the other event $|q_+ \rangle$, then, should have 
an absolute probability equal to 1 to happen. If we consider this case as a 
limiting case when $\theta \rightarrow +\frac{\pi}{4},+\frac{5\pi}{4}$ we 
can see that event $|q_- \rangle$ becomes less and less probable, but the 
corresponding values for the variables ${\widetilde \xi}_{cl}(q_-)$ and 
$h_{cl}(q_-)$ grow towards infinity. In the case of the latter variable 
the probability of the event decreases faster than the divergence of the 
variable, so that this event contributes and infinitesimal amount to the 
average or expected value of the classical value $h_{cl}$ over the set of 
paths. On the contrary, the divergence in the expression for 
the former variable grows fast enough to contribute a finite amount of 
half the value of its total average $<{\widetilde \xi}>$ over paths, even 
though the event has a tiny probability to happen. The situation is 
similar for the event $|q_+ \rangle$ when $e^{i\phi}=\pm 1$ and 
$sin(\theta)=\mp cos(\theta)$.

Let us add that although $\xi_{cl}(t,q_{\pm})$ and 
${\widetilde \xi}_{cl}(t,q_{\pm})$ obey classical equation of an harmonic 
oscillation, the energy of the PCP, as defined by equation (\ref{pce}),
is not necessarily equal to $\frac{1}{2}((\xi_{cl})^2+({\widetilde 
\xi}_{cl})^2)$.

We will finish this section with two results that were already noticed
in \cite{oaknin} but we remind them here for their relevance in the 
discusion of the next sections. The first result:  
\begin{equation}
\label{firstmomentum}
\langle \Psi|O|\Psi \rangle = \langle \Psi|{\cal P}^{\psi}_o(1,\xi)|\Psi 
\rangle = O_{cl}(q_+) |\psi(q_+)|^2 + O_{cl}(q_-) |\psi(q_-)|^2 =
< O_{cl} >_{PCP},
\end{equation}
means not only that the average of pseudoclassical values of the 
observable $O$ over the whole set or ensemble of random PCP's that 
describes the state $|\Psi \rangle$ is equal to the average value of the 
operator $O$ in this state, but it also means that PCP's do not interfere 
between themselves.

The second result on PCP's that we want to bring to attention is related
to the second momentum of the distribution of pseudoclassical values
of the generic operator $O$ over the ensemble of random paths:
\begin{equation}
\label{secondmomentum}
< (O^{\dagger}O)_{cl} >_{PCP} = 
\langle \Psi|O^{\dagger}O|\Psi \rangle = 
\langle \Psi|({\cal P}^{\psi}_o(1,\xi))^{\dagger}
{\cal P}^{\psi}_o(1,\xi)|\Psi \rangle = < O^*_{cl} O_{cl} >_{PCP}.
\end{equation}
It means that the average on PCP's of pseudoclassical values for the
operator $O^{\dagger}O$ is equal to the second momentum of the 
distribution of $O_{cl}$ over the paths. Although this equation holds in 
average for any operator $O$ it does not necessarily hold on each one
of the paths as can be readily seen, for 
instance, from equation (\ref{psi}): while $({\widetilde 
\xi}^{\dagger}{\widetilde \xi})_{cl}=({\widetilde 
\xi}^2)_{cl}=\frac{1}{2}$, we find after some algebra that
$({\widetilde \xi}_{cl})^* ({\widetilde \xi}_{cl}) = \frac{\left(1 +
\frac{1}{2} sin^2(2\theta)\left(1+cos(2\phi)\right)\mp 
2sin(2\theta)\left(cos^2(\theta)cos(\phi)+sin^2(\theta)cos(3\phi)\right)\right)}
{2\left(1-\frac{1}{2}sin^2(2\theta)\left(1+cos(2\phi)\right)\right)}$, 
where the $\mp$ sign in the numerator stands for the value of the random 
variable at $|q_+ \rangle$ and $|q_- \rangle$, respectively. Only after 
averaging these two possible values with their corresponding probabilities 
we recover the identity (\ref{secondmomentum}).
This is the reason why we call these paths pseudoclassical paths: although 
they obey classical equations of motion, the values on paths of physical 
observables which do not commute with the representation of commuting 
observables we have chosen (in this example, the operator $\xi$) do not 
fulfill the usual algebraic relations between classical observables. Or 
stated in other words, $(O^{\dagger}O)_{cl} - (O_{cl})^* (O_{cl})$ is not 
necessarily zero on each one of the paths, neither even on any of them. 
This is a consequence of the non-commutative relations between operators
in the algebra of quantum observables. We will return to discuss this 
aspect of the formalism at the end of next section when we will introduce 
collective observables and show that classical algebraic relations are 
naturally recovered for this kind of operators.

\section{PseudoClassiclal Paths in a system of many fermions}

In this section we discuss how to extend the formalism of 
pseudoclassical paths to systems that contain more than one fermion 
mode, for example a fermionic quantum field. The basic ideas are the same 
that we have already presented in the previous section, but there are now 
some subtleties related to the anticommutation relations between 
operators associated to different fermionic modes that we would like to 
notice. We will start, for the sake of simplicity, with a system with two 
free fermions whose hamiltonian is:
\begin{equation}
\label{hamil2}
H = \kappa_a \left(a^{\dagger}a - \frac{1}{2}\right)+ 
\kappa_b\left(b^{\dagger}b - \frac{1}{2}\right) = H_a + 
H_b.
\end{equation} 
The operators $a^{\dagger}$,$a$ create and annihilate excitations of a 
first fermion mode, and $b^{\dagger}$,$b$ create and annihilate excitations
of a second fermion mode. They obey anticommutation relations
$\{a^{\dagger},a\}=\{b^{\dagger},b\}=1$, and the anticommutator between 
any other pair of these operators is equal to zero. In particular, those
relations imply that $(a^{\dagger})^2=a^2=(b^{\dagger})^2=b^2=0$ and,
moreover, the operations that create or annihilate excitations of the 
first fermionic mode anticommute, instead of commuting, with the operations 
that create or annihilate excitations of the second fermionic mode.

The basis of eigenstates of the hamiltonian (\ref{hamil2}) can be 
obtained applying the creation operators $a^{\dagger}$ and $b^{\dagger}$ 
on the vacuum state $|0;0 \rangle$. The anticommutation relations between 
the operators guarantee that the Fock space they generate contains only 
antisymmetric vectors:
\begin{equation}
\label{twentysix}
|0;0 \rangle \hspace{0.7in}
|1;0 \rangle \equiv a^{\dagger}|0;0 \rangle \hspace{0.7in}
|0;1 \rangle \equiv b^{\dagger}|0;0 \rangle \hspace{0.7in}
|1;1 \rangle \equiv a^{\dagger}b^{\dagger}|0;0 \rangle. 
\end{equation}
Their corresponding eigenvalues can be read of the equalities
\begin{eqnarray}
&H|0;0 \rangle = - \frac{\kappa_a+\kappa_b}{2}|0;0 \rangle \hspace{0.4in}
 H|1;0 \rangle = + \frac{\kappa_a-\kappa_b}{2}|1;0 \rangle& 
\\
&H|0;1 \rangle = - \frac{\kappa_a-\kappa_b}{2}|0;1 \rangle \hspace{0.4in} 
 H|1;1 \rangle = + \frac{\kappa_a+\kappa_b}{2}|1;1 \rangle&. 
\end{eqnarray}

The operators $a^{\dagger},b^{\dagger}$ are the hermitic conjugates of
$a$ and $b$, respectively, so we can define two couples of hermitic 
linear combinations of them: the hermitic operators $\xi$,${\widetilde 
\xi}$ introduced in (\ref{hermitic}), together with the couple 
\begin{equation}
\zeta=\frac{1}{\sqrt{2}}(b+b^{\dagger}), \hspace{1.0in}
{\widetilde \zeta}=\frac{i}{\sqrt{2}}(b-b^{\dagger}).
\end{equation}
These four operators anticommute each other:
\begin{equation}
\label{twofermionanticommutation}
\{\xi,{\widetilde \xi}\}=\{\xi,\zeta\}=\{\xi,{\widetilde \zeta}\}=
\{{\widetilde \xi},\zeta\}=\{{\widetilde \xi},{\widetilde \zeta}\}=
\{\zeta,{\widetilde \zeta}\}=0,
\end{equation}
and each one is proportional to the identity operator when it is 
multiplied by itself, $\xi^2={\widetilde \xi}^2=\zeta^2={\widetilde 
\zeta}^2=\frac{1}{2}$. They generate the Grassmann algebra of linear 
operators defined on the Hilbert space spanned by (\ref{twentysix}).

The two components $H_a$ and $H_b$ of the hamiltonian (\ref{hamil2}) can 
be expressed in terms of the new operators as $H_a=\kappa_a(i{\widetilde 
\xi}\xi)$ and $H_b=\kappa_b(i{\widetilde \zeta}\zeta)$, which commute 
each other $[H_a,H_b]=0$. Therefore,  
$e^{iHt}=e^{i(H_a+H_b)t}=e^{iH_at}e^{iH_bt}$, and
\begin{equation}
\xi(t)=e^{iHt} \xi e^{-iHt}=e^{iH_at} \xi e^{-iH_at},
\end{equation} 
while
\begin{equation}
\zeta(t)=e^{iHt} \zeta e^{-iHt}=e^{iH_bt} \zeta e^{-iH_bt}.
\end{equation} 
Similar expressions are obtained for ${\widetilde \xi}(t)$ and 
${\widetilde \zeta}(t)$ and they imply
\begin{equation}
\label{fermiondynamics1}
\frac{d\xi(t)}{dt} = -\kappa_a{\widetilde \xi}(t),
\hspace{1.0in}
\frac{d{\widetilde \xi}(t)}{dt} = \kappa_a\xi(t)
\end{equation}
\begin{equation}
\label{fermiondynamics2}
\frac{d\zeta(t)}{dt} = -\kappa_b{\widetilde \zeta}(t),
\hspace{1.0in}
\frac{d{\widetilde \zeta}(t)}{dt} = \kappa_b\zeta(t).
\end{equation}

In order to build the set of PCP's which describes a generic
quantum state $|\Psi \rangle$ of this system we need to choose a complete 
representation of commuting observables. The operators 
$\xi$ and $\zeta$ do not commute each other and, therefore, are not 
eligible as such a representation. Instead we use the pair of 
hermitic operators $\xi$ and $i{\widetilde \xi}\zeta$ that, according to 
(\ref{twofermionanticommutation}), do commute each other. 

So we expand the state $|\Psi \rangle=\sum_q \psi(q)|q \rangle$ 
in the basis of common eigenstates $|q \rangle$ to the complete 
representation of commuting observables that we have chosen. We know from 
the discussion of the previous section that each of these eigenstates
is actually promoted to one PCP, whose probability to happen at any time 
is $|\Psi(q)|^2$. Each observable of our representation is naturally 
given as its classical value at each of these eigenstates its own 
eigenvalue on it and then, exploiting identities of the kind $O(t)|\Psi 
\rangle= {\cal P}^{\psi}_o(1,\xi,i{\widetilde \xi}\zeta;t)|\Psi \rangle$, any 
other physical observable $O$ is also assigned a time-dependent value on 
each of them: 
\begin{equation}
O_{cl}(t;q)= {\cal P}^{\psi}_o(t;1,\xi_{cl}(q),(i{\widetilde 
\xi}\zeta)_{cl}(q)).
\end{equation}

We can see from this last expression that whatever the quantum state 
$|\Psi \rangle$ is, equations (\ref{fermiondynamics1}) and 
(\ref{fermiondynamics2}) enforce when they are applied on such state (see 
eq. (\ref{dynamicalequations})) that the two couples of observables 
$\xi$, ${\widetilde \xi}$ and $\zeta$, ${\widetilde \zeta}$ will be 
assigned following this formalism time-dependent values that obey 
classical equations of two decoupled harmonic oscillators,
\begin{equation}
\label{fermiondynamics1classical}
\frac{d\xi_{cl}(t,q)}{dt} = -\kappa_a{\widetilde \xi}_{cl}(t,q),
\hspace{1.0in}
\frac{d{\widetilde \xi}_{cl}(t,q)}{dt} = \kappa_a\xi_{cl}(t,q)
\end{equation}
\begin{equation}
\label{fermiondynamics2classical}
\frac{d\zeta_{cl}(t,q)}{dt} = -\kappa_b{\widetilde \zeta}_{cl}(t,q),
\hspace{1.0in}
\frac{d{\widetilde \zeta}_{cl}(t,q)}{dt} = \kappa_b\zeta_{cl}(t,q),
\end{equation}
on each PCP, generically labeled as $|q \rangle$, of the random set which 
describes that state $|\Psi \rangle$.

As an example we will explicitly construct here the set of PCP's that 
describes the vacuum state $|0;0 \rangle$ of this system. The 
generalization to other states is straightforward following the 
three-steps algorithm discussed in the previous section and we will only 
outline how to apply it to the two-fermions system. 

The first stage then is to find the way to express the action on the 
vacuum state $|0;0 \rangle$ of any operator $O$ in the Grassmann algebra 
generated by $\xi$,${\widetilde \xi}$, $\zeta$ and ${\widetilde \zeta}$ 
in terms of the action on the same state of some linear combination of 
the set of operators generated by the complete representation of 
commuting observables, $\xi$ and $i{\widetilde \xi}\zeta$:
\begin{equation}
\label{twins}
O|0;0 \rangle = (\alpha_1 \cdot 1 + \alpha_2 \cdot \xi + \alpha_3 \cdot 
(i{\widetilde \xi} \zeta) + \alpha_4 \cdot \xi(i{\widetilde 
\xi}\zeta))|0;0 \rangle \equiv {\cal P}_o(1,\xi,i{\widetilde 
\xi}\zeta)|0;0 \rangle.
\end{equation}
The operator ${\cal P}_o(1,\xi,i{\widetilde \xi}\zeta)$ 
defined by this expression is unique because the action of the 
identity operator $1$ on the vacuum does not excite any mode, the action 
of the operator $\xi$ excites the mode $|1;0 \rangle$, the action of the 
operator $(i{\widetilde \xi}\zeta)$ excites $|1;1 \rangle$ and the action 
of $(i\xi{\widetilde \xi}\zeta)$ excites $|0;1 \rangle$. Therefore, the 
only linear combination of these four operators which can give the number 
zero when acting on the vacuum is the trivial combination. 

For example, we already know that ${\widetilde 
\xi}|0;0 \rangle=-i\xi|0;0 
\rangle$ because of the identity $a|0;0 \rangle=0$. 
A little more tricky is the way we describe the action 
of the operator $\zeta$ on the vacuum. We first note that 
$\zeta=2{\widetilde \xi}({\widetilde \xi}\zeta)=
-2({\widetilde \xi}\zeta){\widetilde \xi}$.
Therefore, $\zeta|0;0 \rangle=-2({\widetilde 
\xi}\zeta){\widetilde \xi}|0;0 \rangle$, and subtituting in 
this expression the previous result we get
$\zeta|0;0 \rangle=2(i{\widetilde 
\xi}\zeta)\xi|0;0 \rangle=2\xi(i{\widetilde 
\xi}\zeta)|0;0 \rangle$. 
Now, using the identity $b|0;0 \rangle=0$, we get
${\widetilde \zeta}|0;0 \rangle=-i\zeta|0;0 
\rangle=-2i\xi(i{\widetilde \xi}\zeta)|0;0 \rangle$. Using the 
anticommutation relations between the generators of the Grassmann algebra 
we can write down similar expression for any other operator $O$. 

The next stage is to obtain the basis of common eigenvectors to the 
complete representation of commuting observables we have chosen,$\xi$ and 
$(i{\widetilde \xi}\zeta)$: 
\begin{equation}
\label{2pp}
|q_+;q_+ \rangle \equiv \frac{1}{2}\left(  
(|0;0 \rangle+|1;0 \rangle)+(|0;1 \rangle+|1;1 \rangle)\right)
\end{equation}
\begin{equation}
\label{2pm}
|q_+;q_- \rangle \equiv \frac{1}{2}\left(  
(|0;0 \rangle+|1;0 \rangle)-(|0;1 \rangle+|1;1 \rangle)\right)
\end{equation}
\begin{equation}
\label{2mp}
|q_-;q_+ \rangle \equiv \frac{1}{2}\left(  
(|0;0 \rangle-|1;0 \rangle)-(|0;1 \rangle-|1;1 \rangle)\right)
\end{equation}
\begin{equation}
\label{2mm}
|q_-;q_- \rangle \equiv \frac{1}{2}\left(  
(|0;0 \rangle-|1;0 \rangle)+(|0;1 \rangle-|1;1 \rangle)\right),
\end{equation}
and their corresponding eigenvalues, 
\begin{equation}
\label{pp}
\xi|q_+;q_+ \rangle=\frac{1}{\sqrt{2}}|q_+;q_+ \rangle
\hspace{1.4in}
i{\widetilde \xi}\zeta|q_+;q_+ \rangle=+\frac{1}{2}|q_+;q_+ 
\rangle
\end{equation}
\begin{equation}
\label{pm}
\xi|q_+;q_- \rangle=\frac{1}{\sqrt{2}}|q_+;q_- \rangle
\hspace{1.4in}
i{\widetilde \xi}\zeta|q_+;q_- \rangle=-\frac{1}{2}|q_+;q_- 
\rangle
\end{equation}
\begin{equation}
\label{mp}
\xi|q_-;q_+ \rangle=-\frac{1}{\sqrt{2}}|q_-;q_+ \rangle
\hspace{1.4in}
i{\widetilde \xi}\zeta|q_-;q_+ \rangle=+\frac{1}{2}|q_-;q_+ 
\rangle
\end{equation}
\begin{equation}
\label{mm}
\xi|q_-;q_- \rangle=-\frac{1}{\sqrt{2}}|q_-;q_- \rangle
\hspace{1.4in}
i{\widetilde \xi}\zeta|q_-;q_- \rangle=-\frac{1}{2}|q_-;q_- 
\rangle.
\end{equation}

In the new basis (\ref{2pp})-(\ref{2mm}) the vacuum state can be expanded 
as
\begin{equation}
|0;0 \rangle=\frac{1}{2}\left(
|q_+;q_+ \rangle + |q_+;q_- \rangle + |q_-;q_+ \rangle + |q_-;q_- \rangle
\right) \equiv \sum_{q^a_{\pm},q^b_{\pm}} 
\psi(q^a_{\pm},q^b_{\pm})|q^a_{\pm};
q^b_{\pm} \rangle.
\end{equation}
The configuration space of this system has four disconnected points, each 
one of them has, when the system is in its vacuum, a probability 
$|\psi(q)|^2=\left(\frac{1}{2}\right)^2$ to happen randomly.

The classical values at each point of the configuration space 
of the operators that form our complete representation of commuting 
observables can be obtained directly from (\ref{pp})-(\ref{mm}). For example, 
$\xi_{cl}(q_+,q_+)=\frac{\langle q_+;q_+|\xi|0;0 
\rangle}{\langle q_+;q_+|0;0 \rangle}=\frac{1}{\sqrt{2}}$ and 
$(i{\widetilde \xi}\zeta)_{cl}(q_+;q_+)=\frac{\langle q_+;q_+|
i{\widetilde \xi}\zeta|0;0 \rangle}{\langle q_+;q_+|0;0 
\rangle}=\frac{1}{2}$. Similarly, we define their classical values in 
the other three points of the configuration space.

Now we can benefit from identity (\ref{twins}) in order to assign 
to any other operator $O$ its classical value at each point of the 
configuration space: $O_{cl}(q_{\pm},q_{\pm})=
{\cal P}_o(1,\xi_{cl}(q_{\pm},q_{\pm}),(i{\widetilde 
\xi}\zeta)_{cl}(q_{\pm},q_{\pm}))=
\frac{\langle q_{\pm};q_{\pm}|O|0;0 \rangle}{\langle q_{\pm};q_{\pm}|0;0 
\rangle}$. For example, ${\widetilde 
\xi}_{cl}(q_{\pm};q_{\pm})=-i\xi_{cl}(q_{\pm};q_{\pm})$.
Of course, $(i\xi{\widetilde \xi}\zeta)_{cl}(q_{\pm};q_{\pm})=
\xi_{cl}(q_{\pm};q_{\pm})\cdot(i{\widetilde 
\xi}\zeta)_{cl}(q_{\pm};q_{\pm})$, because the two operators of our 
representation commute each other. Then, $\zeta_{cl}(q_{\pm};q_{\pm})
=2(i{\widetilde \xi}\zeta)_{cl}(q_{\pm};q_{\pm})\cdot
\xi_{cl}(q_{\pm};q_{\pm})$ and
${\widetilde \zeta}_{cl}(q_{\pm};q_{\pm})=-i\zeta_{cl}(q_{\pm};q_{\pm})$.
The dependence on time of these random variables can now be obtained from
the equations of motion (\ref{fermiondynamics1classical}), 
(\ref{fermiondynamics2classical}).

Once we have stated the rules to define the random variables on 
configuration space associated to each operator we can pick $\xi_{cl}$ and
$\zeta_{cl}$ as the two independent variables, noticing then that
$(i{\widetilde \xi}\zeta)_{cl}=\frac{1}{2}\zeta_{cl}/\xi_{cl}$. In the
vacuum the four equally probable random events in the configuration space 
correspond to the four possibilities: $\xi_{cl}=\pm \frac{1}{\sqrt{2}}$,
$\zeta_{cl}=\pm \frac{1}{\sqrt{2}}$.

The whole formalism can be repeated for any quantum state $|\Psi 
\rangle$ other than the vacuum. First, we need to expand the state
$|\Psi \rangle$ in the basis (\ref{2pp})-(\ref{2mm}): 
$|\Psi \rangle=\psi(+;+)|q_+;q_+ \rangle+\psi(+;-)|q_+;q_- 
\rangle+\psi(-;+)|q_-;q_+ \rangle+\psi(-;-)|q_-;q_- 
\rangle$. Each PCP correspond to one of the points $|q^a_{\pm};q^b_{\pm} 
\rangle$ in the configuration space and its probability to happen 
randomly is equal to the modulus squared of the amplitude of the 
wavefunction, $|\langle q|\Psi \rangle|^2$. The four events are no 
longer necessarily equally probable, but the classical 
values on each of them of the random variables assigned to the operators 
that form our complete representation of commuting observables are still 
given by their corresponding eigenvalues (\ref{pp})-(\ref{mm}). 
Then, the action of the operator $O(t)$ on the state $|\Psi \rangle$ is 
written as the action on the vacuum of some operator ${\cal 
G}(t;1,\xi,i{\widetilde \xi}\zeta)$: $O(t)|\Psi \rangle= (\chi_{0;0}(t) 
\cdot 1 + \chi_{1;0}(t) \cdot \xi + \chi_{1;1}(t) \cdot (i{\widetilde \xi} 
\zeta) + \chi_{0;1}(t) \cdot \xi(i{\widetilde \xi}\zeta))|0;0 \rangle 
\equiv {\cal G}(t;1,\xi,i{\widetilde \xi}\zeta)|0;0 \rangle$. This step is 
always possible and uniquely defined because the action on the vacuum of 
each one of the four operators $1$,$\xi$,$(i{\widetilde \xi} \zeta)$ and 
$\xi(i{\widetilde \xi}\zeta))$ excites a different mode in the basis 
(\ref{twentysix}) of eigenstates of the hamiltonian (\ref{hamil2}). 
Finally, the relation between the states $|\Psi \rangle$ and $|0;0 
\rangle$ is inverted $|0;0 \rangle= {\cal J}(1,\xi,i{\widetilde 
\xi}\zeta)|\Psi \rangle$ and introduced in the previous expression to get
the desired result: $O(t)|\Psi \rangle={\cal G}(t;1,\xi,i{\widetilde 
\xi}\zeta) {\cal J}(1,\xi,i{\widetilde \xi}\zeta)|\Psi \rangle= 
{\cal P}^{\psi}_o(t;1,\xi,i{\widetilde \xi}\zeta)|\Psi \rangle$. This last
expression can be used to define the classical value of the operator
$O$ in $|q \rangle$ when the quantum system is described in its state
$|\Psi \rangle$ as $O_{cl}(t;q)={\cal P}^{\psi}_o(t;1,\xi_{cl}(q),(i{\widetilde 
\xi}\zeta)_{cl}(q))$. 

We see that although each PCP corresponds to one of the eigenstates 
$|q_{\pm};q_{\pm} \rangle$, the way how we actually define the classical 
values of physical observables explicitly depends on the quantum state of 
the system $|\Psi \rangle$. Therefore, the PCP cannot be identified with 
the quantum state described by the eigenstate $|q \rangle$. Instead the 
whole set of four random PCP's, which according to 
(\ref{fermiondynamics1})-(\ref{fermiondynamics2}) 
should still obey classical equation of motion, does describe the 
dynamics of the system in its quantum state $|\Psi \rangle$. The state
of the system, therefore, enters this description by: first, fixing the 
different probabilities of each of the four points of the configuration 
space; second, defining the initial conditions on the set of classical 
equations (\ref{fermiondynamics1})-(\ref{fermiondynamics2}) for each of 
the PCP's; third, defining the algebraic relations between the values of 
different physical observables on each PCP. 

The formalism can now be straightforward generalized to a fermionic system
containing many modes. We define a couple of hermitic operators 
$\xi_{\it l}$,${\widetilde \xi}_{\it l}$ associated to each one of these 
modes. The integer index ${\it l}=0,1,2,...$ labels 
the different modes. The set of operators $\xi_0$, $i{\widetilde \xi}_0\xi_1$,  
$i{\widetilde \xi}_1\xi_2$,... forms a complete representation of 
commuting observables and the Hilbert space of the states of the system 
can be linearly spanned in the basis of common eigenstates 
to all the operators in this representation. Each common eigenstate can be 
labeled by a sequence of $+$ or $-$ signs (for example,  
$|+,+,-,+,-,-,-,...\rangle$) each input corresponding to one 
of the fermionic modes. In total, $2^{\cal N}$ different eigenstates, 
where ${\cal N}$ is the number of different modes. A $+$ sign in the 
zeroth position corresponds to an eigenstate with eigenvalue 
$\frac{+1}{\sqrt{2}}$ for the operator $\xi_0$. Inversely, a $-$ sign 
correspond to an state with eigenvalue $\frac{-1}{\sqrt{2}}$ for this 
operator. And similarly, a $\pm$ sign in the ${\it l}$-th position 
corresponds to an state with eigenvalue $\frac{\pm 1}{2}$ for the operator 
$i{\widetilde \xi}_{{\it l}-1}\xi_{{\it l}}$, ${\it l} \ge 1$. 
In the quantum state $|\Psi \rangle = \sum_{q^{(n)}_{\pm}} \psi(q^{(n)}_{\pm}) 
|q^{(n)}_{\pm} \rangle$ each of the $2^{\cal N}$ random pseudoclassical 
paths $|q \rangle$ which describe the system has a probability 
$|\psi(q)|^2$ to happen. In the vacuum all the paths are equally probable 
and, therefore, this probability is $1/2^{\cal N}$. 

Let us now turn back to equations (\ref{firstmomentum}) and 
(\ref{secondmomentum}), which describe the statistical behaviour of the 
random variables over the ensemble of PCP's. They relate the first 
and second momentum of the random variable $O_{cl}$ to the expected values
of the operator $O$ and $O^{\dagger}O$, respectively. The proof of these 
equations that we gave in the previous section also proves the assertions 
for systems with many fermionic modes. We noticed then that  
(\ref{secondmomentum}) holds only on average over 
the whole ensemble of PCP's but does not necessarily hold on each 
specific PCP and understood that, as a consequence of the 
non-commutative relations between quantum operators, the usual algebraic 
relations between classical observables are not necessarily fulfilled by 
the time-dependent values $O_{cl}(t)$ defined on the non-interfering paths. 
There are, nevertheless, a certain class of operators, which we call 
"collective" operators for reasons that will become clear in the next 
section, for which $(O^{\dagger} O)_{cl} - (O_{cl})^* O_{cl} \sim 0$ not 
only in the average sense of equation (\ref{secondmomentum}), but under 
the stronger requirement that
\begin{equation}
\label{collective}
\sigma_O \equiv \left[ <\left((O^{\dagger} O)_{cl} - (O_{cl})^* 
O_{cl}\right)^2>_{PCP} \right]^{1/2} \hspace{0.3in} 
\ll \hspace{0.3in} <(O_{cl})^* O_{cl}>_{PCP}. 
\end{equation}
In this sense $(O^{\dagger} O)_{cl} \sim (O_{cl})^* O_{cl}$ on each PCP 
independently of the quantum state $|\Psi \rangle$ that they describe 
and we can conclude that values on paths of collective observables 
approximately recover the usual algebraic relations between classical 
observables. This condition is naturally satisfied for some operators 
which depend on the dynamics of many modes because for these operators the 
random variable $(O^{\dagger} O)_{cl} - (O_{cl})^* O_{cl}$ can be 
expressed as a quadratic function of many other independent random 
variables associated to individual modes and, according to the central 
limit theorem, condition (\ref{collective}) is satisfied. Moreover, higher 
statistical moments of the collective random variable are highly
suppressed. 

\section{PCP's in fermionic field theories}

We are now ready to develop the formalism of PCP's for the Dirac theory
of free spinors. We consider a single fermionic field $\psi(x)$
in Minkowski space-time, whose lagrangian density is \footnote{we use 
units in which $\hbar=1$, $c=1$.}
\begin{equation}
\label{lagrangian}
{\cal L} = \bar{\psi}\left(i\partial_{\mu}\gamma^{\mu}-M \right)\psi,
\end{equation}
where $\gamma^{\mu}$ are $4$D Clifford matrices and $M$ is the mass of the 
spinor field. The hamiltonian of the system is:
\begin{equation}
\label{hamiltonianF}
{\cal H} = \int d^3{\vec x} 
\hspace{0.1in}
\bar{\psi}({\vec x})\left( -i \gamma^{j} 
\partial_{j} + M \right)\psi({\vec x}).
\end{equation}

In order to regularize the theory in the infrared limit we impose
periodic boundary conditions on the large three-dimensional box
$\left[0,X\right] \times \left[0,X\right] \times \left[0,X\right]$.
The spinor field $\psi({\vec x})$ and its conjugate $\bar{\psi}({\vec 
x}) \equiv \psi^{\dagger}({\vec x})\gamma^0$ can then be expanded in a 
discrete series of Fourier modes: 
\begin{equation}
\label{field}
\psi({\vec x}) = \sum_{n_1,n_2,n_3=-N}^{+N} \frac{1}{X^{3/2}} \sum_s
\left(a^s_{\vec n} u^s_{\vec n} e^{+2\pi i{\vec n}\cdot{\vec x}/X} +
b^{s\dagger}_{\vec n} v^s_{\vec n} e^{-2\pi i {\vec n}\cdot{\vec x}/X} 
\right),
\end{equation}
\begin{equation}
\label{fieldbar}
\bar{\psi}({\vec x}) = \sum_{n_1,n_2,n_3=-N}^{+N} \frac{1}{X^{3/2}} \sum_s
\left(a^{s\dagger}_{\vec n} \bar{u}^s_{\vec n} e^{-2\pi i{\vec 
n}\cdot{\vec x}/X} + b^{s}_{\vec n} \bar{v}^s_{\vec n} e^{+2\pi i {\vec 
n}\cdot{\vec x}/X}\right).
\end{equation}

The index $s$ takes two possible values which correspond to two
possible helicity orientations for the fermionic mode labeled by the index 
${\vec n} \equiv (n_1,n_2,n_3)$. The infinite series of Fourier modes 
has been cut by the ultraviolet regulator $N/X$. The operators 
$a^s_{\vec n}$, $b^s_{\vec n}$ and their hermitic conjugates  
$a^{s\dagger}_{\vec n}$, $b^{s\dagger}_{\vec n}$ obey canonical 
anticommutation relation: 
$\{a^s_{\vec n},a^{s'\dagger}_{\vec m}\}=\{b^s_{\vec 
n},b^{s'\dagger}_{\vec m}\}=\delta^{s s'} \delta_{{\vec n}{\vec m}}$ and 
any other anticommutator between these operators is equal to zero.
They create and annihilate quanta of the fermion and its antifermion, 
respectively. The Dirac spinors $u^{s=\pm}_{\vec n}$ are two
linearly independent solutions to the equation 
$\left[- i\gamma^j\partial_j + M\right]
u^s_{\vec n} e^{+2\pi i{\vec n}\cdot {\vec x}/X} =
\kappa_{\vec n} \gamma^0 u^s_{\vec n} e^{+2\pi i{\vec n}\cdot {\vec x}/X}$,
while $v^{s=\pm}_{\vec n}$ solve the equation
$\left[- i\gamma^j\partial_j + M\right]
v^s_{\vec n} e^{-2\pi i{\vec n}\cdot {\vec x}/X} =
-\kappa_{\vec n} \gamma^0 v^s_{\vec n} e^{-2\pi i{\vec n}\cdot {\vec x}/X}$,
where $\kappa_{\vec n}=\left[|\frac{2\pi{\vec n}}{X}|^2+M^2 
\right]^{1/2}$. They obey orthogonality relations of the kind
$u^{s\dagger}_{\vec n} v^{s'}_{-{\vec n}} = v^{s\dagger}_{\vec n} 
u^{s'}_{-{\vec n}} = 0$ and are normalized such that 
$u^{s\dagger}_{\vec n} u^{s'}_{\vec n} = v^{s\dagger}_{\vec n} 
v^{s'}_{\vec n} = \delta^{s s'}$.

Once we put (\ref{field}) and (\ref{fieldbar}) into (\ref{hamiltonianF}) 
we obtain the following expression for the hamiltonian of the system
\begin{equation}
\label{hamiltonian0}
{\cal H} = \sum_{{\vec n},s} \kappa_{\vec n} \left( 
(a^{s\dagger}_{\vec n} a^{s}_{\vec n}-\frac{1}{2}) + 
(b^{s\dagger}_{\vec n} b^{s}_{\vec n}-\frac{1}{2}) \right), 
\end{equation}
which describes a collection of free fermionic modes. 

We introduce for each mode, labeled by the pair of indexes $s,{\vec n}$, a 
couple of hermitic operators $\xi^s_{\vec n}$ and ${\widetilde 
\xi}^s_{\vec n}$ as we did in the previous section. The task of choosing 
a complete representation of commuting observables is much simplified 
after having regularized the theory in the IR and UV: then we have a 
large but finite number of fermionic modes that we can order in a 
sequence. The set of hermitic operators $\xi_0$, $i{\widetilde \xi}_0 
\xi_1$,$i{\widetilde \xi}_1 \xi_2$, ...forms a complete representation
of commuting observables and we can expand any quantum state $|\Psi \rangle$ 
of the system in the basis of common eigenstates $|q \rangle$ to the 
operators in this representation.

Following the formalism that we have developed in the two previous 
sections we can give to each physical observable $O({\vec x},t)$ a 
"classical" value $O_{cl}({\vec x},t)$ at each of these common eigenstates 
$|q \rangle$, which are so promoted to describe space-time dependent 
pseudoclassical field configurations (PCFC's) whose probabilities to
randomly happen are, again, given by $|\langle q|\Psi \rangle|^2$. We 
remark again that although each PCFC corresponds to a certain eigenstate 
$|q \rangle$, the way how we actually assign classical values to the 
physical observables, exploiting identities like (\ref{ground}), 
explicitly depends on the quantum state $|\Psi \rangle$ of the system 
and, therefore, the path cannot be identified with the quantum state 
described by $|q \rangle$.

The state $|\Psi \rangle$ of the quantum field is then 
represented as a linear combination of randomly distributed PCFC's. Each 
field operator, in particular $\psi(\vec x)$ and its conjugate 
$\bar{\psi}({\vec x})$, is 
assigned a time-dependent value on each of these paths. The operator equation
\begin{equation}
\label{operator equation}
\left[i \partial_{\mu}\gamma^{\mu} - M\right]\psi(t,{\vec x}) = 0,
\end{equation}
and its hermitic conjugate
\begin{equation}
\label{operator antiequation}
\bar{\psi}(t,{\vec x})\left[i \partial_{\mu}\gamma^{\mu} + M\right] = 0,
\end{equation}
when are applied on the quantum state of the system
\begin{equation}
\label{operator equationPsi}
\left[i \partial_{\mu}\gamma^{\mu} - M\right]\psi(t,{\vec x})
|\Psi \rangle = 0, \hspace{0.7in}
\bar{\psi}(t,{\vec x})\left[i \partial_{\mu}\gamma^{\mu} + M\right]
|\Psi \rangle = 0,
\end{equation}
guarantee that on each of the random PCFC's in the set that describes the
quantum state the classical values $\psi(t,{\vec x})_{cl}$ and 
$\bar{\psi}(t,{\vec x})_{cl}$  obey classical equations of motion:
\begin{equation}
\label{operator equationclassical}
\left[i \partial_{\mu}\gamma^{\mu} - M\right]\psi(t,{\vec x})_{cl} = 0,
\hspace{0.7in}
\bar{\psi}(t,{\vec x})_{cl}\left[i \partial_{\mu}\gamma^{\mu} + M\right] 
= 0.
\end{equation}
On the other hand, the initial conditions on each of the PCFC in 
the set that describes the state, as well as their probabilities 
to randomly happen, are different for each quantum state of the system, 
as we have discussed in the previous sections. Let us remark that 
$(\bar{\psi})_{cl}=(\psi^{\dagger}\gamma^0)_{cl}$ 
is determined from the action of the operator on the ket-state 
$\bar{\psi}|\Psi \rangle$ and, 
therefore, is not equal to $(\psi_{cl})^{\dagger}\gamma^0$ which results 
from its action on the bra-state $\langle \Psi|\bar{\psi}$. 

We can now consider the addition of a linear interaction term to the 
quantum field free theory that we are discussing. If the coupling is weak 
enough we can work in perturbation theory. Then, the Hilbert space of states
can still be built as the Fock space of free quanta and the algebra of 
the quantum operators is not altered. So, the whole formalism of PCFC's
can be repeated. Now there is a new interacting term that appears in 
the hamiltonian of the system and modify the dynamics of the quantum 
operators, $O(t)=e^{i(H_0+\lambda H_{int})t} O e^{-i(H_0+\lambda 
H_{int})t}$, in the Heisenberg picture. In consequence, it also modifies
the time-dependent values of the field operators on each PCFC.
We can see from the operator equations (\ref{operator equation}) and 
(\ref{operator antiequation}) that $\psi(t,{\vec x})_{cl}$ and  
$\bar{\psi}(t,{\vec x})_{cl}$ will still obey exactly classical equations 
of motion because the term added to the equations is linear in the 
operators $\psi(t,{\vec x})$ and $\bar{\psi}(t,{\vec x})$.

As a simple example, we can add a Yukawa coupling term to the lagrangian
(\ref{lagrangian}). This interaction adds a new term to the operator
equation:
\begin{equation}
\label{operator equationWyukawa}
\left[i \partial_{\mu}\gamma^{\mu} - M\right]\psi(t,{\vec x}) +
g \phi(t,{\vec x}) \psi(t,{\vec x}) = 0.
\end{equation}
This equation implies, when it is applied on the quantum state $|\Psi 
\rangle$, that the dynamics of the pseudoclassical values on its
PCFC's gets modified: 
\begin{equation}
\label{operator equationWyukawaclass}
\left[i \partial_{\mu}\gamma^{\mu} - M\right]\psi(t,{\vec x})_{cl} +
g \phi(t,{\vec x})_{cl} \psi(t,{\vec x})_{cl} = 0
\end{equation}
and a similar equation can be obtained for $(\bar{\psi})_{cl}$.
In the new term we have made use of the identity
$(\phi \psi)_{cl}=(\phi)_{cl} (\psi)_{cl}$, which can be easily proved 
because in a weakly interacting theory the operators $\phi$ and $\psi$ act 
on different sectors of the Hilbert space, and have taken advantage of
the linearity of (\ref{operator equationWyukawa}) in the fermion field
operator. The interaction term does not modify the initial conditions to 
be fixed on the PCFC's that describe the quantum state, which depend only 
on the relations between operators when they act on the quantum state. 
Neither it does modify the random probability of each path. A discussion 
of, in general, non-linear interactions will be presented in a forthcoming 
work.

In the picture that we have presented quantum states of fermionic fields
are represented as linear combination of randomly distributed paths that
obey classical Dirac equations and do not interfere between themselves.
The formalism is reminiscent of the consistent histories approach of 
Gell-Mann and Hartle \cite{GMH}, wherein the Feynman paths 
\cite{Feynmann} which contribute to a certain process are grouped into
equivalence classes or non-detailed histories that are approximately
incoherent. On the contrary, in the formalism that we have developed the 
random paths do not interfere at all and obey "deterministic" equations 
of motion at the price of redefining the algebraic relations between 
the values on paths of physical observables.

We can use this formalism of PCFC's to describe quantum fluctuations of 
fermionic fields in terms of pseudoclassical stochastic proccesses. In 
particular, we can use it to describe local fluctuations of globally 
conserved numbers using classical concepts. We present here two examples 
that could be interesting: local fluctuations of the energy and fermion 
number in a finite volume $V$ contained in the whole box 
$\left[0,X\right]^3$ and parametrically larger than the UV cutoff that we 
have introduced.  

The operator ${\cal H}_V = \int_V d^3{\vec x} \hspace{0.1in} 
\bar{\psi}({\vec x})\left(-i \gamma^{j} \partial_{j} + 
M \right)\psi({\vec x})$ which describes the energy contained in the 
finite volume $V$ was already introduced in \cite{eichler} 
and discussed in \cite{oaknin} in the context of a bosonic field theory. 
Similarly the fermion number contained in the same volume is 
described by the operator ${\cal B}_V = \int_V d^3{\vec x} \hspace{0.1in} 
\bar{\psi}({\vec x})\gamma^0 \psi({\vec x})$. 

If we want to find the "classical" expressions for these two operators in 
the set of PCFC's that describe a generic quantum field theoretic state 
$|\Psi \rangle$ we would need to write the action on this state of each of 
the operators in terms of the action on the same state of some function of  
the operators in a complete representation of communting observables. We 
have done this excersise in previous sections and, therefore, we will 
present here a different approach that will shortcut the way towards this 
aim and will give us a different perspective on PCFC's.

We know from equation (\ref{secondmomentum}) that
\begin{eqnarray*}
< ({\cal H}_V(t))_{cl} >_{PCFC} 
& = & \int_V d^3{\vec x} \hspace{0.1in} 
< (\bar{\psi}(t,{\vec x})\left(-i \gamma^{j} \partial_{j} + 
M \right)\psi(t,{\vec x}))_{cl} >_{PCFC}  \\
& = &\int_V d^3{\vec x} \hspace{0.1in} 
< (\psi_{cl}(t,{\vec x}))^{\dagger}\gamma^0\left(-i \gamma^{j} 
\partial_{j} + 
M \right)\psi_{cl}(t,{\vec x}) >_{PCFC}, 
\end{eqnarray*}
and similarly
\begin{eqnarray*}
< ({\cal B}_V(t))_{cl} >_{PCFC} 
& = & \int_V d^3{\vec x} \hspace{0.1in} 
< (\bar{\psi}(t,{\vec x}) \gamma^0 \psi(t,{\vec x}))_{cl} >_{PCFC}  \\
& = &\int_V d^3{\vec x} \hspace{0.1in} 
< (\psi_{cl}(t,{\vec x}))^{\dagger} \psi_{cl}(t,{\vec x}) 
>_{PCFC}. 
\end{eqnarray*}
Although these equations hold only on average over the whole ensemble of 
PCFC's and not necessarily on each of them, we gave in (\ref{collective}) 
a condition that ensures that
\begin{equation}
\label{classicalenergy}
({\cal H}_V(t))_{cl} \simeq \int_V d^3{\vec x} \hspace{0.1in} 
(\psi_{cl}(t,{\vec x}))^{\dagger}\gamma^0\left(-i \gamma^{j} \partial_{j} 
+ M \right)\psi_{cl}(t,{\vec x}), 
\end{equation}
and
\begin{equation}
\label{classicalnumber}
({\cal B}_V(t))_{cl} \simeq \int_V d^3{\vec x} \hspace{0.1in} 
(\psi_{cl}(t,{\vec x}))^{\dagger} \psi_{cl}(t,{\vec x}), 
\end{equation}
on each PCFC, if the random variables
$\left[({\cal H}_V)_{cl} - \int_V d^3{\vec x} \hspace{0.1in} 
(\psi_{cl}({\vec x}))^{\dagger}\gamma^0\left(-i \gamma^{j} \partial_{j} + 
M \right)\psi_{cl}({\vec x})\right]$ and
$\left[({\cal B}_V)_{cl} - \int_V d^3{\vec x} \hspace{0.1in} 
(\psi_{cl}({\vec x}))^{\dagger} \psi_{cl}({\vec x})\right]$
are quadratic in many independent random variables. We called this 
kind of observables collective observables and will show below that these 
two operators are in fact "collective" operators, but for 
now let us just assume it to get to some conclussions:

a) We can describe the dynamics on paths of "collective" observables like 
(\ref{classicalenergy}) and (\ref{classicalnumber}) in terms of 
classical stochastic concepts, from the classical 
Dirac equations (\ref{operator equationclassical}) or (\ref{operator 
equationWyukawaclass}) that $\psi_{cl}(t,{\vec x})$ obeys. Only the 
initial conditions and the actual distribution of paths are constrained by 
the quantum state of the system $|\Psi \rangle$. 

b) Then, the time correlation function of any operator can be defined 
on each PCFC as usual in statistichal mechanics \cite{landau}, 
$f_o(t_1-t_2)=(O_{cl}(t_1))^* O_{cl}(t_2)-(O_{cl}(t_2))^* O_{cl}(t_1)$, 
and the lifetime of the fluctuation is defined as the inverse width of its 
Fourier transform. We proved in \cite{oaknin} that the average over the
ensemble of PCFC's coincides, as it should do, with the usual definition 
of the time scale of quantum fluctuations.

We will conclude this section giving a formal proof
that justifies (\ref{classicalenergy}) and (\ref{classicalnumber}). Let
start introducing the Fourier expansions (\ref{field}) and 
(\ref{fieldbar}) in the expressions for the operators ${\cal H}_V$ and 
${\cal B}_V$. Their classical values on paths can then be expressed as
\footnote{We have included all the dependence on the geometry of the volume
$V$ inside the spatial integrals $F({\vec k},X)=\int_V d^3x \frac{1}{X^3} 
exp\left(-2\pi i{\vec k}\cdot{\vec x}/X\right)$, that can be analytically
computed for simple geometries. If $V=\left[0,X\right]^3$ is the whole
box, $F({\vec k},X)=0$ except for $F({\vec k}=0,X)=1$.}:
\begin{equation}
\label{hamiltonianV}
({\cal H}_V)_{cl} =
\sum_{{\vec n},{\vec m}} 
\sum_{s,s'} \kappa_{\vec m} 
\left[(a^{s\dagger}_{\vec n} \bar{u}^s_{\vec n} +
b^{s}_{-{\vec n}} \bar{v}^s_{-{\vec n}})
(a^{s'}_{\vec m} \gamma^0 u^{s'}_{\vec m} -
b^{s'\dagger}_{-{\vec m}} \gamma^0 v^{s'}_{-{\vec m}})\right]_{cl}
F({\vec n}-{\vec m},X),
\end{equation}
\begin{equation}
\label{fermionnumberV}
({\cal B}_V)_{cl} = \sum_{{\vec n},{\vec m}} 
\sum_{s,s'} 
\left[(a^{s\dagger}_{\vec n} \bar{u}^s_{\vec n} +
b^{s}_{-{\vec n}} \bar{v}^s_{-{\vec n}})
(a^{s'}_{\vec m}\gamma^0 u^{s'}_{\vec m} +
b^{s'\dagger}_{-{\vec m}}\gamma^0 v^{s'}_{-{\vec m}})\right]_{cl}
F({\vec n}-{\vec m},X).
\end{equation}
On the other hand the approximate expressions 
(\ref{classicalenergy}),(\ref{classicalnumber}) can be expressed
using the same Fourier expansions as:
\begin{equation}
\label{hamiltonianVaprox}
({\cal H}_V)^{\it{approx}}_{cl} =
\sum_{{\vec n},{\vec m}} 
\sum_{s,s'} \kappa_{\vec m} 
\left[([a^{s}_{\vec n}]^*_{cl} \bar{u}^s_{\vec n} +
[b^{s^{\dagger}}_{-{\vec n}}]^*_{cl} \bar{v}^s_{-{\vec n}})
([a^{s'}_{\vec m}]_{cl} \gamma^0 u^{s'}_{\vec m} -
[b^{s'\dagger}_{-{\vec m}}]_{cl} \gamma^0 v^{s'}_{-{\vec m}})\right]
F({\vec n}-{\vec m},X),
\end{equation}
\begin{equation}
\label{fermionnumberVaprox}
({\cal B}_V)^{\it{approx}}_{cl} = \sum_{{\vec n},{\vec m}} 
\sum_{s,s'} 
\left[([a^{s}_{\vec n}]^*_{cl} \bar{u}^s_{\vec n} +
[b^{s\dagger}_{-{\vec n}}]^*_{cl} \bar{v}^s_{-{\vec n}})
([a^{s'}_{\vec m}]_{cl}\gamma^0 u^{s'}_{\vec m} +
[b^{s'\dagger}_{-{\vec m}}]_{cl}\gamma^0 v^{s'}_{-{\vec m}})\right]
F({\vec n}-{\vec m},X).
\end{equation}
We need to prove that $({\cal H}_V)_{cl}-({\cal H}_V)^{\it{approx}}_{cl}$
and $({\cal B}_V)_{cl}-({\cal B}_V)^{\it{approx}}_{cl}$ are bilinear in
the random variables associated to the single modes. Comparing both 
expressions we can see that it is enough to prove that 
expressions of the kind $[a^{s\dagger}_{\vec n}a^{s'}_{\vec m}]_{cl}-
[a^{s}_{\vec n}]^*_{cl}[a^{s'}_{\vec m}]_{cl}$, which involve only two 
different fermion modes are bilinear. We know from (\ref{twins})
that they are.

\section{Summary}

We have extended to systems of fermions the formalism of 
pseudoclassical paths that we recently developed for 
systems of weakly interacting bosons and have shown that fermionic quantum 
states can also be represented in the Heisenberg picture as linear 
combinations of randomly distributed paths which do not interfere 
between themselves. Every physical observable was assigned a 
time-dependent value on each path in a way that respects the 
anticommutation relations between fermionic operators and, in 
consequence, these values on paths do not neccesarily respect the usual 
algebraic relations between classical observables. Each path depicts a 
collection of pseudoclassical harmonic oscillators with 
constrained initial conditions. 

We used these paths to define the dynamics of quantum 
fluctuations in systems of fermions without reference to an 
environment or any additional external system. Then, we selected 
"collective" observables which depend on many fermionic degrees of 
freedom and found that, as we found in \cite{oaknin} for collective 
operators which depend on many bosonic modes, their values on PCP's do 
approximately fulfill the usual algebraic relations between classical 
observables and, therefore, the dynamics of these collective fluctuations 
can be described in terms of unconstrained classical stochastic processes. 

In this setup, we showed that quantum fluctuations of fermionic fields 
obey classical Dirac equations and described the dynamics of 
local fluctuations of globally conserved fermion numbers.

\section{Acknowledgments}

I thank R. Brustein for his help and for encouraging me to write this 
paper, which broaden the ideas already presented in \cite{oaknin}. 
I warmly acknowledge many useful discussions with J. Oaknin . This work 
was partly supported by the National Science and Engineering Research 
Council of Canada.


\begin{references}

\bibitem{Hartle:nn}
J.~B.~Hartle and B.~L.~Hu,
Phys.\ Rev.\ D {\bf 21}, 2756 (1980).

\bibitem{reynaud}
M.~T.~Jaekel and S.~Reynaud,
Rept.\ Prog.\ Phys.\  {\bf 60}, 863 (1997)
[arXiv:quant-ph/9706035].

\bibitem{linde}
L.~A.~Kofman and A.~D.~Linde,
Nucl.\ Phys.\ B {\bf 282}, 555 (1987).

\bibitem{polarski}
D.~Polarski and A.~A.~Starobinsky,
Class.\ Quant.\ Grav.\  {\bf 13}, 377 (1996)
[arXiv:gr-qc/9504030].

\bibitem{garriga}
J.~Garriga and T.~Tanaka,
Phys.\ Rev.\ D {\bf 65}, 103506 (2002) 
[arXiv:hep-th/0112028].
\bibitem{bucher}
M.~Bucher,
arXiv:hep-th/0107148.
J.~J.~Blanco-Pillado and M.~Bucher,
Phys.\ Rev.\ D {\bf 65}, 083517 (2002)
[arXiv:hep-th/0111089].

\bibitem{negele} J.W. Negele and H. Orland, {\it Quantum Many-Particle 
Systems}. Redwood City, California: Addison-Wesley Pub. Co., (1998);
A. Galindo, {\it Quantum Mechanics}, Vol. 1 (Text and 
Monographs in Physics). Springer-Verlag, (1991).

\bibitem{oaknin}  R. Brustein and D.H. Oaknin, 
Phys.\ Rev.\ D {\bf 67}, 025010 (2003)
[arXiv:hep-th/0207251].

\bibitem{Halliwell:1998jf}
J.~J.~Halliwell,
Phys.\ Rev.\ D {\bf 58}, 105015 (1998)
[arXiv:quant-ph/9805062].

\bibitem{GMH}
M.~Gell-Mann and J.~B.~Hartle,
Phys.\ Rev.\ D {\bf 47}, 3345 (1993) [arXiv:gr-qc/9210010].

\bibitem{Feynmann} R. Feynmann and A. Hibbs, {\it Quantum
Mechanics and Path Integrals}, McGraw-Hill Book Co, New York (1965).

\bibitem{eichler} R. Brustein, D. Eichler, S. Foffa and D.H. Oaknin,
Phys.\ Rev.\ D {\bf 65}, 105013 (2002) 
[arXiv:hep-th/0009063].

\bibitem{landau} L.D. Landau and E.M. Lifshitz, {\it Statistical
Mechanics}, Vol. 1. Oxford:Pergamon Press, (1969)

\end{references}
\end{document}